# The diagnostic utility of endocytoscopy for the detection of esophageal lesions: a systematic review and meta-analysis

[1]Lu wang[1];Bofu Tang [1,*];Feifei Liu[2];Zhenyu Jiang[3,&];Xianmei Meng[4,&]

【 Abstract 】 **Objective** To systematically evaluate the value of endocytoscopy (ECS) in the diagnosis of early esophageal cancer (EC). **Methods** Pubmed, Ovid and EMbase databases were searched to collect diagnostic tests of ECS assisted diagnosis of early EC. The retrieval time was from the establishment of the database to August 2022. Review manager 5.4, Stata 16.0 and Meta-Disc 1.4 were used for meta-analysis after two researchers independently screened literature, extracted data and evaluated the bias risk of included studies. **Results** A total of 7 studies were included, including 520 lesions. Meta-analysis results showed that the combined sensitivity(SE), specificity(SP), positive likelihood ratio (PLR), negative likelihood ratio (NLR), diagnostic odds ratio (DOR) and positive posterior probability (PPP) of ECS screening for early EC were 0.95[95%*CI*: 0.84, 0.98], 0.92 [95%*CI*: 0.83, 0.96], 11.8 [95%*CI*: 5.3, 26.1], 0.06 [95%*CI*: 0.02, 0.18], 203 [95%*CI*: 50, 816], and 75%, respectively. The area (AUC) under the summary receiver Operating Characteristic curve (SROC) was 0.98[95%*CI*: 0.96, 0.99]. **Conclusions** Current evidence suggests that ECS can be used as an effective screening tool for early EC. Due to the limited number and quality of included studies, it is imperative to conduct more high-quality studies to verify the above conclusions.

---

[1] *Lu Wang and Bofu Tang are co-first authors;
&Zhenyu Jiang and Xianmei Meng are Co-corresponding authors.

**Key words**：Esophageal cancer(EC); Endocytoscopy（ECS）; Diagnosis; Meta-analysis

## 1. Background

EC is a malignant tumor originating from the esophageal mucosal epithelium of the digestive tract. It is known as the seventh most common cancer and the sixth most fatal cancer type worldwide, seriously threatening people's life and health. [1,2] EC was often diagnosed at an advanced stage, with high mortality and poor prognosis . Therefore, the diagnosis and treatment of early EC are quite important. Early EC lesions are small, lacking characteristics, so the diagnosis of early EC necessitates high diagnostic performance of endoscopists and endoscopic equipments, as well as multiple biopsies of suspected lesions [3,4] . In view of this, it is imperative to explore new auxiliary diagnostic methods to improve the diagnosis of early EC.

Digestive endoscopy has been widely used in the diagnosis of various esophageal diseases. In recent years, with the continuous improvement of medical devices for detecting early EC, international attention has been intensely focused on realizing the endoscopic observation at the cell level, on gradually reducing the dependence on invasive pathological biopsy for recognizing and interpreting mucosal lesions, and on improving the diagnostic efficiency of cytology. ECS, a novel magnification endoscope with high precision, can observe tiny blood vessels, acinar cavity structure and nuclear morphology, thus, having been popular in clinical practice recently. [5-7]

Though the accuracy of ECS in diagnosing of EC has been reported, the results of different studies vary greatly. [8,9] The paper conducted a meta-analysis of the studies regarding the diagnosis of EC via ECS, in order to understand the diagnostic value of ECS for EC, hoping to provide clinical guidance.

2. Methods

2.1 Retrieval of articles

A comprehensive search was conducted on Pubmed, Ovid and EMbase databases published up to July 2022, using the following search terms: "endocytoscopy", "endocytoscopic", "ECS", "esophageal cancer" and "esophagus cancer". Search for published articles on the diagnosis of EC by ECS was conducted to obtain comprehensive data. All articles were written in English.

## 2.2 Selection of Articles

**2.2.1** The publications included in this systematic review met the following inclusion criteria: ① the objective was to evaluate the accuracy of ECS in diagnosing EC; ② all articles were published and available ; ③ a prospective or retrospective study design was conducted ; ④ adult participants were included; ⑤ all articles included the following information: area under curve (AUC), sensitivity(SE), specificity(SP), positive likelihood ratio (PLR), negative likelihood ratio (NLR), diagnostic odds ratio (DOR), or accuracy, directly or indirectly offering true positive(TP), false positive(FP), false negative(FN), and true negative (TN); ⑥ ECS and pathological biopsy were performed in all studies; and ⑦ the reference standard was pathological biopsy.

**2.2.2** The publications included in this systematic review met the following exclusion criteria: ① the examination methods that were combined with acetic acid staining, indigo carmine staining or fecal occult blood test; ② narrative reviews, comments, proceedings, or study protocols；and ③ results in vitro.

## 2.3 Assessment of methodological quality

The methodological quality of the final articles was assessed by two researchers using the second version of Quality Assessment of Diagnostic

Accuracy Studies (QUADAS-2) tool.[10] This tool comprises 4 domains, including "patient selection," "index test," "reference standard," and "flow and timing", with the first three featuring an "applicability" assessment. Each was evaluated as "high risk", "low risk", or "unclear risk" of bias by two researchers. If all relevant questions in this part are evaluated as "Yes", then this part is judged as "low risk". If the answer to one or more of the questions is "no", this part is rated as "high risk"; Others were judged "unclear". Two researchers completed this part of the work independently, and any disagreement was resolved through discussion.

## 2.4 Extraction of data

Data were extracted independently by two researchers, including author, country, year, number of lesions, reference standard and type of study.

## 2.5 Methods of statistics

Statistical analysis of the data was performed using Stata 16, Meta-Disc (Version 1.4) and Review manager 5.4 software. Review manager 5.4 was used to fill in the content included in the studies and draw quality assessment pictures according to QUADAS-2 standard. Spearman correlation coefficient and summary receiver operating characteristic curve (SROC) were used to determine whether threshold effect existed.

Between-study heterogeneity was determined using the Cochran's Q test and $I^2$ statistic. A bivariate mixed-effects binary regression model was applied to determine pooled effect estimates. SE, SP, PLR, NLR, DOR, AUC and 95%CI were calculated. Deeks' funnel plot and test were used to assess the publication bias. $P < 0.05$ was considered statistically significant.

## 3. Results

### 3.1 Results of literature selection

A total of 443 English articles were retrieved. After 135 duplicate articles were deleted, 308 articles remained. Based on the titles and abstracts, 289 articles were excluded. After the full text of the articles was obtained and read carefully, 12 articles were excluded. Finally, 7 articles were included, including 520 lesions. The literature search and screening process is shown in Figure 1. The baseline characteristics of the literature are shown in Table 1.

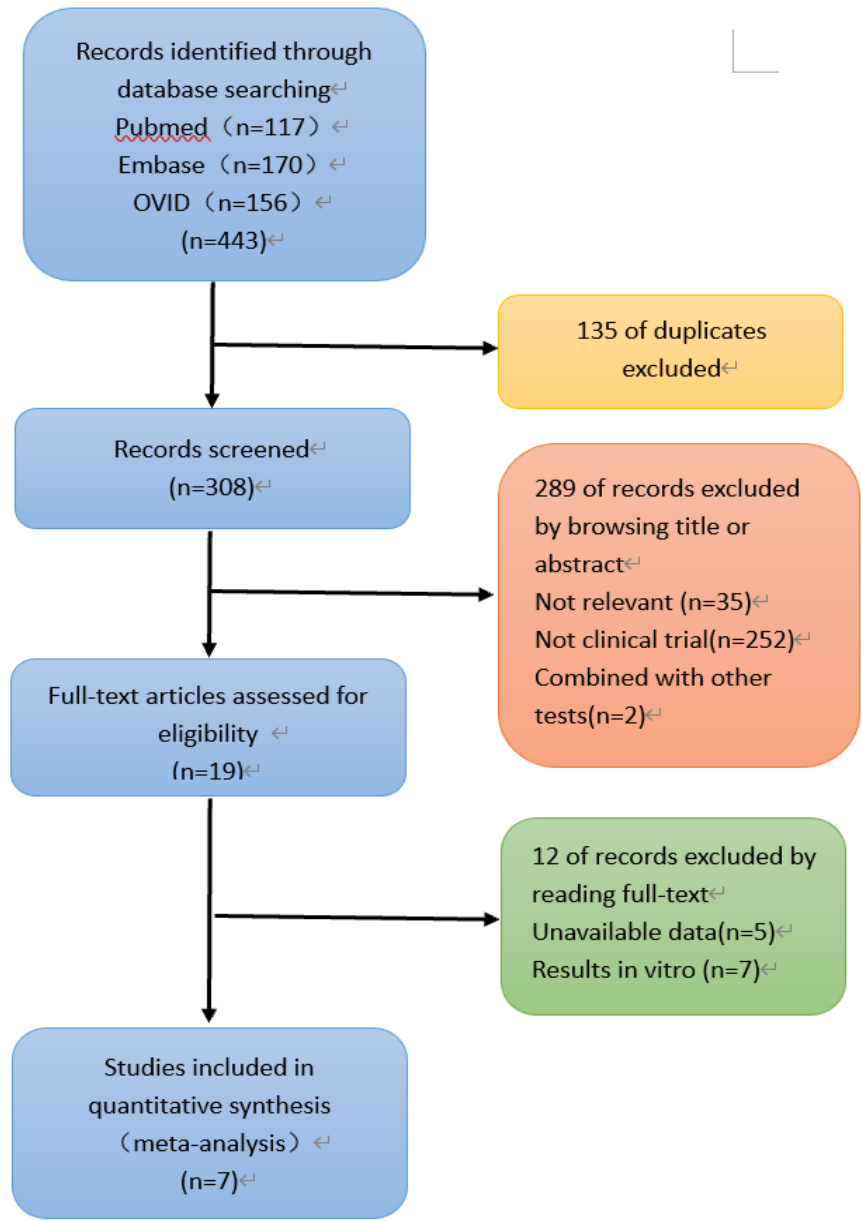

Figure 1 Flow diagram for study selection.

## 3.2 Results of literature quality evaluation

The data of the 7 articles [9, 11-16] included in the study were completed. In accordance with the QUADAS-2 standard, the quality evaluation and mapping of all the included literatures were conducted by using RevMan 5.4 software. The specific quality evaluation results of the literature were shown in Figure 2.

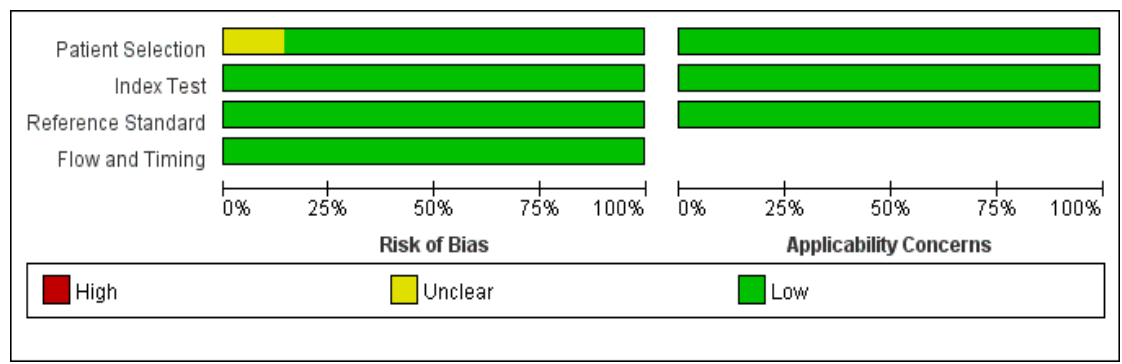

Note: Risk of Bias: deviation risk; Applicability Concerns: clinical applicability; Patient Selection: patient selection; Index Test: test to be evaluated; Reference Standard: Gold standard; Flow and Timing: Case flow and progress

Figure 2 Quality evaluation of the articles

**3.3 Heterogeneity Analysis**

Heterogeneity tests showed that $I^2 = 66.73\%$ ($P = 0.01$) for sensitivity, $I^2 = 72.43\%$ ($P < 0.0001$) for specificity, $I^2 = 54.25\%$ ($P < 0.0001$) for PLR, $I^2 = 59.03\%$ ($P = 0.02$) for NLR, and $I^2 = 97.42\%$ ($P < 0.0001$) for DOR, which suggests the presence of heterogeneity was unrelated to threshold effects in this study, the source of heterogeneity was analyzed. Due to the limited number (＜10) of included articles, meta-analysis was not conducted to analyze the source of heterogeneity. The Galbraith radial plot shows no heterogeneity. (Figure 3)

### 3.4 Evaluation of publication bias

The Deeks' test was performed using Stata software to assess publication bias. As shown in Figure 4, P = 0.77 suggests that there was no significant publication bias in the included studies.

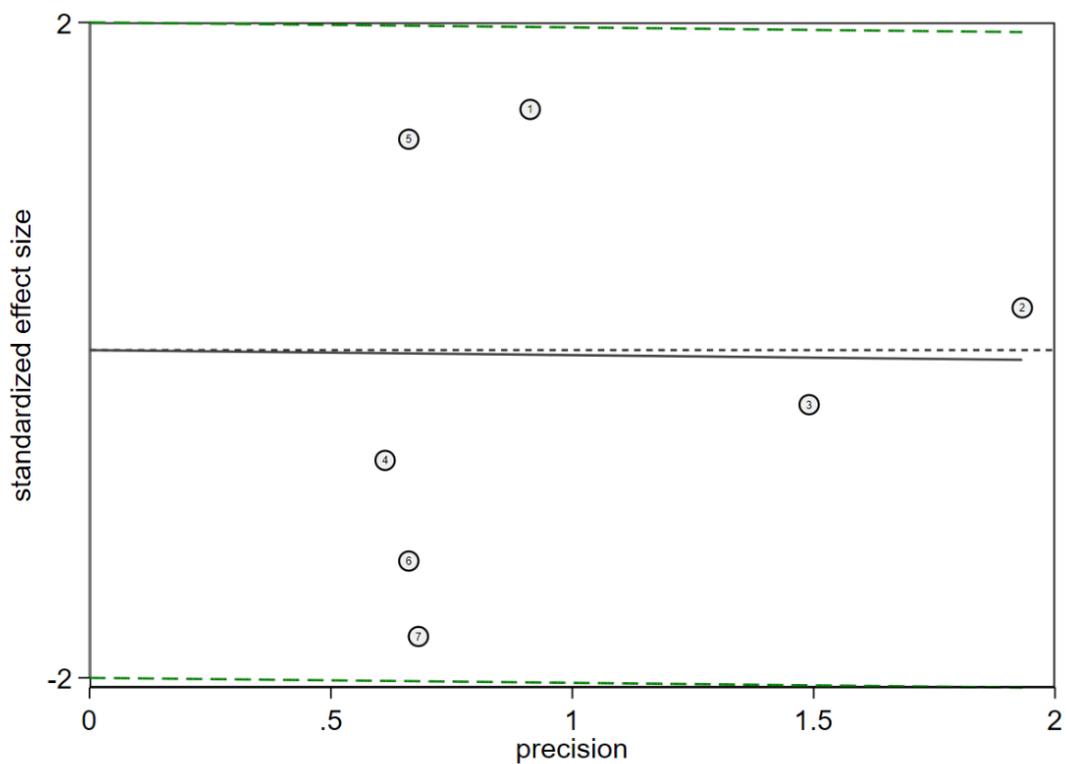

Figure 3 Heterogeneity analysis. Heterogeneity was evaluated by Galbraith radial plot.

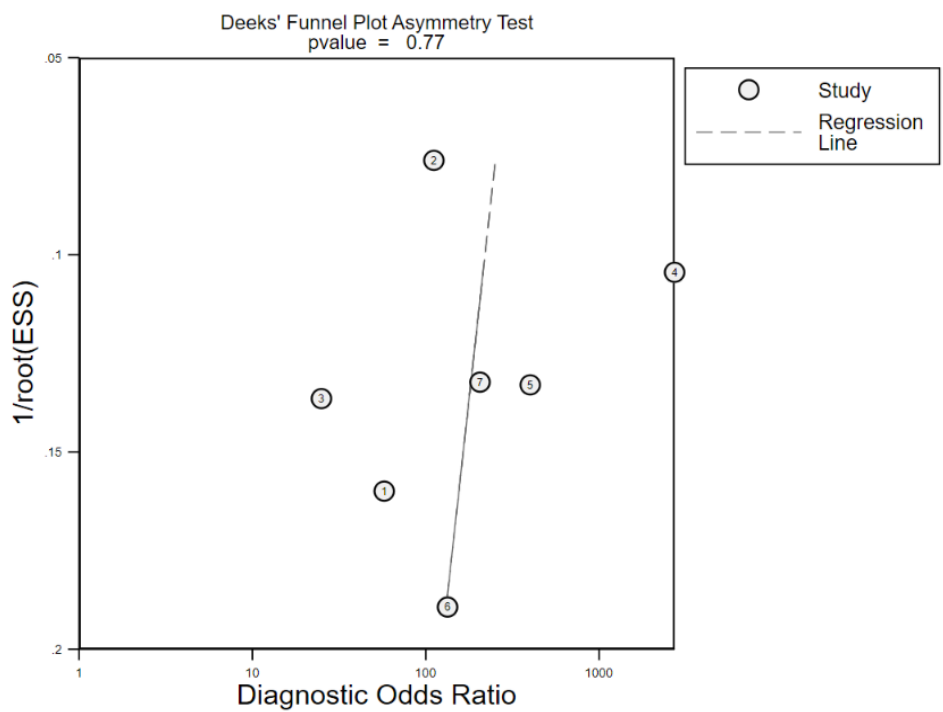

Figure 4　　Deek's Funnel Plot Asymmetry Test

### 3.5 Results of meta-analysis

Results of meta-analysis on the accuracy of ECS in the diagnosing early EC: First, heterogeneity test was performed on the 7 included articles, and the Spearman correlation coefficient of threshold effect was 0.286 (P = 0.535), suggesting no heterogeneity caused by threshold effect. Second, a coupled forest plot of SE and SP was generated, and the combined SE, SP,PLR,NLR, and diagnostic odds ratio of ECS in the diagnosis of early EC were 0.95 [95%*CI*：0.84, 0.98], 0.92 [95%*CI*：0.83, 0.96], 11.8 [ 95%*CI*：5.3, 26.1 ] ,0.06 [ 95%*CI*：0.02, 0.18 ], 203 [95%*CI*：50, 816], respectively . Third, the SROC curve was generated for the diagnosis of EC with ECS, and the total AUC was 0.98[ 95%*CI*：0.96, 0.99 ]. Fourth, a scatter plot of positive and negative likelihood ratios with combined summary point is created, shows that PLR is 12, NLR is 0.06, and the aggregate point of PLR and NLR was located in the upper left quadrant. Finally, the Fagan graph showed that the probability of ECS previously classified as early EC increased from a positive average diagnosis rate of 20% to 75%, while the probability of reduction was 1% when it was negative. (Figure 5-Figure 8)

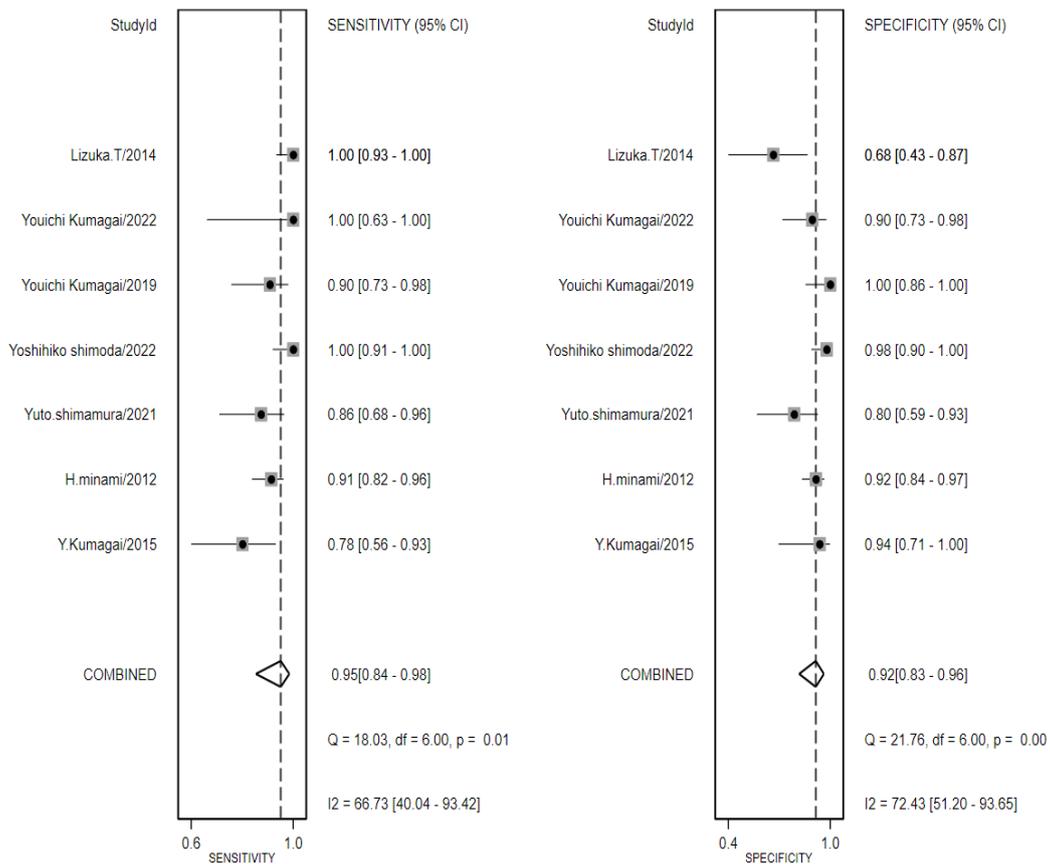

Figure 5 The combined SE and SP of ECS for the diagnosis of early EC.

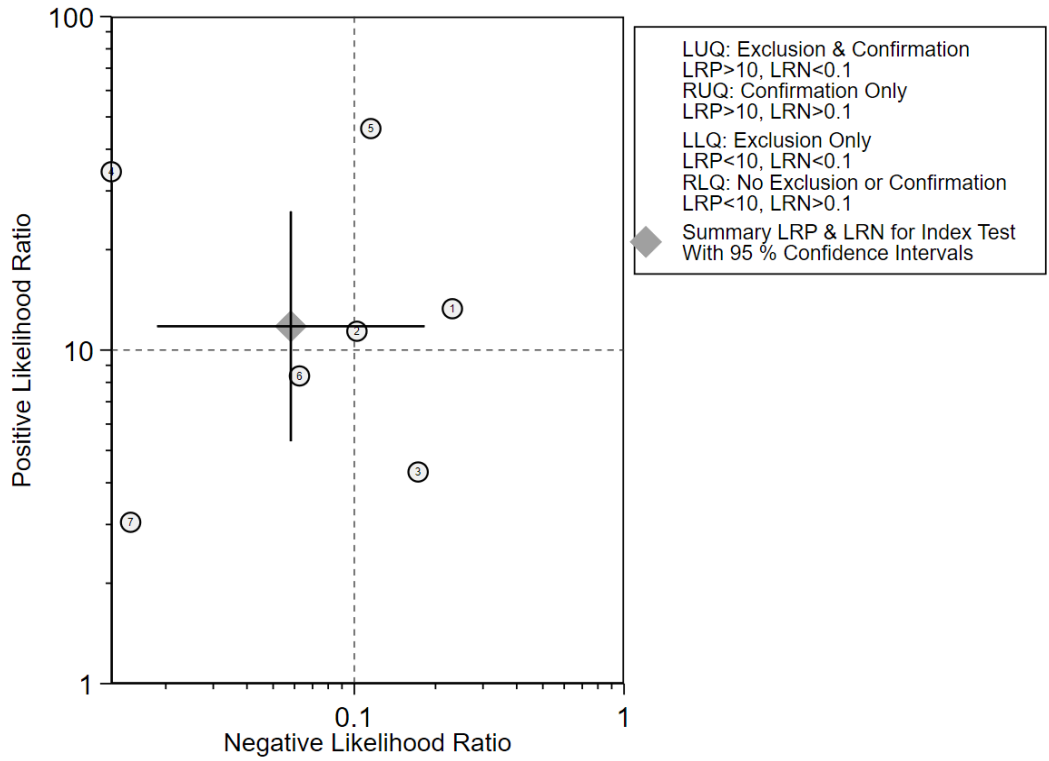

Figure 6　The likelihood ratio scatter plot

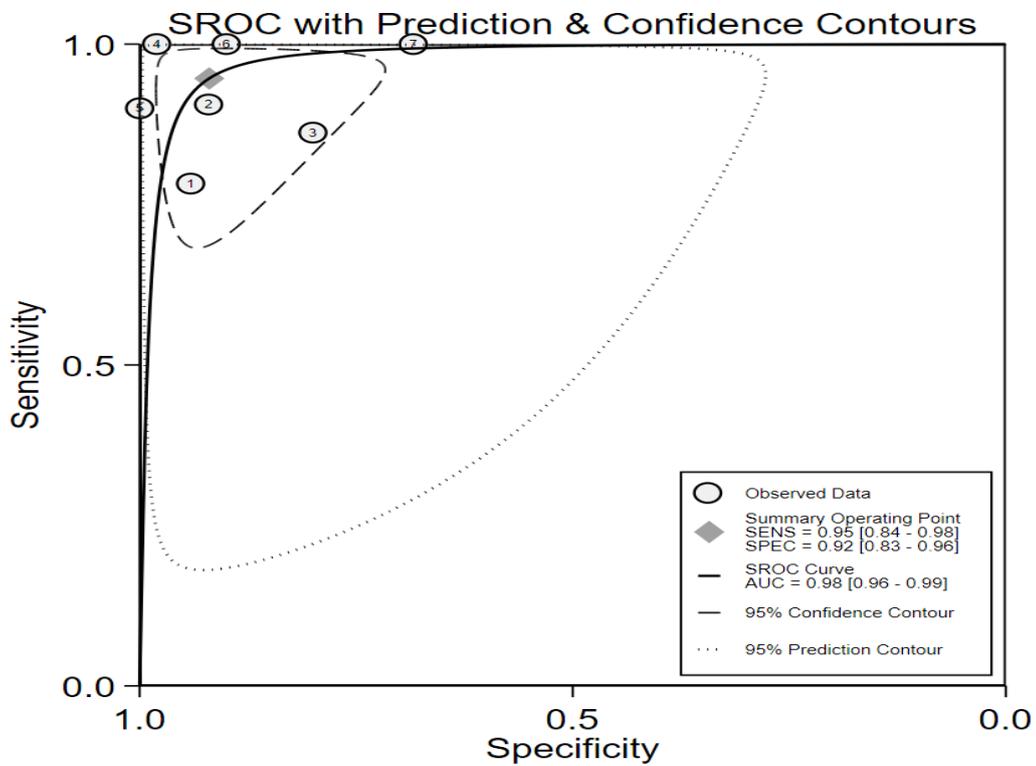

Note: Summary receiver operating characteristic (SROC) curve with 95% confidence region and prediction region of ECS diagnosis　for the diagnosis of early EC . AUC,

area under the curve; SE, sensitivity; SP, specificity

Figure 7　　SROC curve

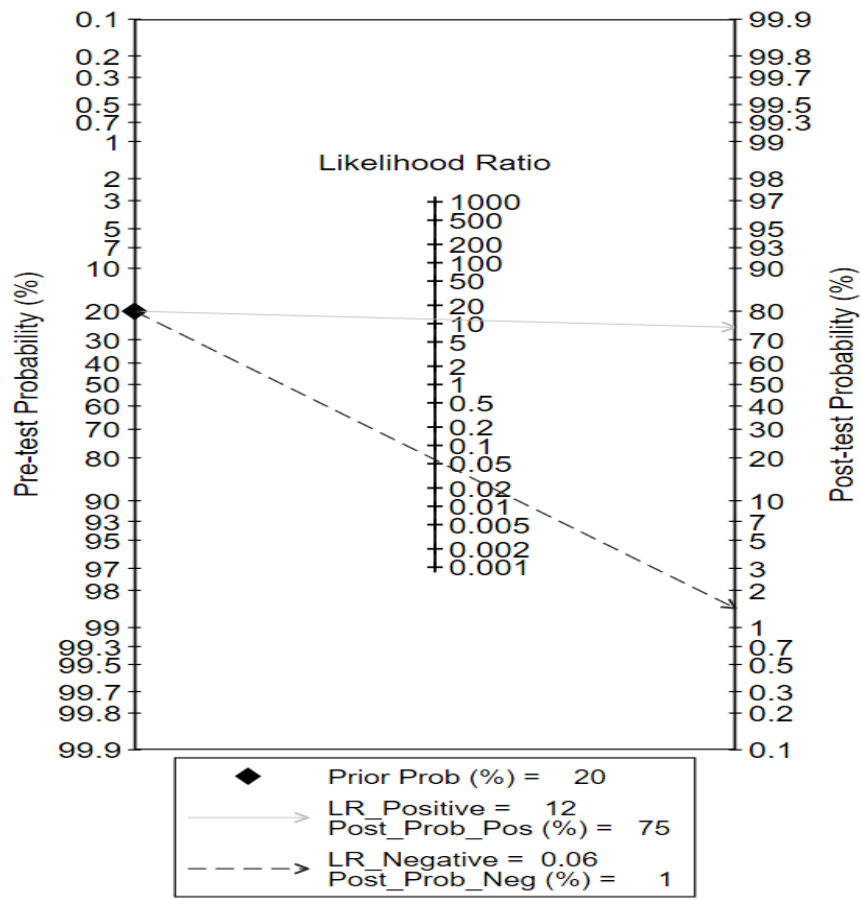

Figure 8　　Fagan diagram

## 4. Discussion

Due to inadequate physical examination and poor early diagnosis, EC remains one of the major threats to human health. To reduce mortality and improve prognosis, early diagnosis and treatment of EC is an effective way.[17] Currently, endoscopic pathological biopsy is efficient in early diagnosing EC, but this method requires endoscopic clinicians to possess rich surgical and pathological experience. In addition, biopsy is ineffective in detecting subtle lesions, and non-standard operation during surgery may pose a risk of bleeding and injury of esophageal mucosa. In order to overcome these challenges, an increasing number of diagnostic researches focus on exploring tumor biomarkers and developing optical equipment.

The widely used fourth generation of ECS is a novel ultra-high magnification endoscopic technique, featuring 500× continuous zoom magnification and an observation range of 570 μm × 500 μm.[18] ECS, combined with narrow-band imaging technology and double dye staining (crystal violet and methylene blue), allows endoscopists to judge the morphology of esophageal lesions by naked eye, thus realizing real-time biopsy in vivo characterized by simple operation, little trauma, high sensitivity and specificity.[19] The endocytoscopic diagnosis for esophageal neoplasms has been actively investigated because the first research on EC was performed for the esophagus .[20,21] The comparision of ECS and conventional pathological analysis can be seen in the Table 2.

|  | ECS | Conventional Pathological |
|---|---|---|
| **Staining before examination** | √ | × |
| **Post-biopsy bleeding** | × | √ |
| **Fibrosis of the post-biopsy scar** | × | √ |
| **Pathology requirements for endoscopists** | √ | × |

Notes: " √" in the table means needed or exists, and "×" in the table means not needed or will not happen.

Table 1 Comparision of ECS and conventional pathological analysis

In this study, PLR and NLR for diagnosis of early EC were 12 and 0.06, respectively. It is generally believed that the chances of diagnosing or ruling out the disease greatly increase when PLR>10 or NLR<0.1. The AUC of the SROC curve is 0.98 (AUC＞0.9 indicates higher accuracy). The Fagan chart suggests that ECS can better diagnose early EC: If a patient is defined as having a prior probability of 20%, then the PPP is 75% and negative posterior probability (NPP) is 1%. That is, when the probability of the patient suffering from early EC is judged to be 20% according to the symptoms and signs, the patient has a 75% probability of being diagnosed with the disease if the ECS test result is positive, while this probability is 1% if the ECS test results are negative. Together, the above results indicate that ECS is highly valuable in the diagnosis of early EC.

The image classification systems of the ECS have not been standardized, there are "three categories", "four categories" and "five categories" according to the morphology of nuclei after staining. Among them, the modified three-level classification method is the most commonly used.[22,23] The difference in image classification systems is one of the important reasons for the high heterogeneity in the paper. In addition, high heterogeneity may occur for the following reasons: (1) different studies used varying types of ECS; and (2) without the assistance of pathologists, endoscopists make different judgments based on ECS images. The difference in the accuracy of the diagnosis of lesions by endoscopists also restrict the clinical application of ECS.

In recent years, artificial intelligence (AI), especially as a medical image screening system, has made remarkable progress in varied medical fields. More importantly, AI will be an effective way to reduce the above-mentioned heterogeneity. Youichi Kumagai et al [16,23] found that compared with endoscopists, AI-assisted ECS had significantly high overall accuracy, sensitivity and specificity in judging the same esophageal lesions, reaching 90.9%, 92.6% and 89.3%, respectively. In 2022, the diagnostic accuracy, sensitivity and specificity of ECS were 94.7%, 91% and 96%, respectively, using an improved AI specificity. The AI systems showed high diagnostic performance that was equal or superior to that of experienced endoscopists. With the diagnostic accuracy similar to that of pathologists，it is expected to become an effective aid for ECS examination in the future.

Limitations and shortcomings of this study: the studies included in this analysis were incomplete, and few prospective randomized blind trials specifically investigated the diagnostic effect of ECS in diagnosing early EC. In order to improve the quality of research, it is suggested that the design of clinical trials should be more rigorous. Meta-analysis cannot replace a large number of clinical study samples, thus requiring more clinical studies for validation. Due to the limitations of the included study design, multiple biases are unavoidable. Although the funnel plot indicates a small possibility of publication bias, the accuracy of the research results may be affected because (1) all included studies stemmed from Japan without grey literature being retrieved; and (2) the included studies presented certain heterogeneity, the sources of which could not be explored due to the limited number of included studies and data.

## 5. Conclusion

ECS have been shown to have good diagnostic accuracy, as well as the advantages of being less invasive and easy to operate, and thus can be used as an effective tool for diagnosing esophageal lesions. The application of ECS in the diagnosing gastrointestinal tumors, especially early EC, remains a challenge. However, its accuracy will be further improved with the development of endoscopy equipment and AI intelligence, as well as accumulation of clinical experience.

| Study | Year | Nationality | Study format | Type of ECS | Total number of lesions | ECA Classification | Reference Standard | TP | FP | FN | TN | Performance of ESC with AI (TP/FP/FN/TN) |
|---|---|---|---|---|---|---|---|---|---|---|---|---|
| Y.Kumagai[11] | 2015 | Japan | Retrospective | GIF-Y0002 | 40 | Four categories | Histological diagnosis | 18 | 1 | 5 | 16 | -- |
| H.minami[12] | 2012 | Japan | Retrospective | GIF-Y0001/GIF-Y0002 | 173 | Five categories | Histological diagnosis | 77 | 7 | 8 | 81 | -- |
| Yuto.shimamura[13] | 2021 | Japan | Retrospective | GIF-Y0074/GIF-H290EC | 55 | Three categories | Histological diagnosis | 25 | 5 | 4 | 20 | -- |
| Yoshihiko shimoda[14] | 2022 | Japan | Retrospective | GIF-H290EC | 91 | Three categories | Histological diagnosis | 40 | 1 | 0 | 50 | -- |
| Youichi Kumagai[15] | 2019 | Japan | Retrospective | GIF-Y0002/GIF-Y0074 | 55 | Three categories | Histological diagnosis | 27 | 0 | 3 | 25 | 25/2/3/25 |
| Youichi Kumagai[16] | 2022 | Japan | Retrospective | GIF-H290EC | 38 | Three categories | Histological diagnosis | 8 | 3 | 0 | 27 | 10/1/1/26 |
| Lizuka.T[9] | 2014 | Japan | Retrospective | Not mentioned | 68 | Five categories | Histological diagnosis | 49 | 6 | 0 | 13 | -- |

Table 1　Baseline characteristics of the literature